\input harvmac.tex


\def\ap{{\alpha^\prime}}
\def\myTitle#1#2{\nopagenumbers\abstractfont\hsize=\hstitle\rightline{#1}%
\vskip 0.5in\centerline{\titlefont #2}\abstractfont\vskip .5in\pageno=0}

%
%
\def\np#1#2#3{Nucl. Phys. {\bf B#1} (#2) #3}
\def\pl#1#2#3{Phys. Lett. {\bf #1B} (#2) #3}
\def\plb#1#2#3{Phys. Lett. {\bf #1B} (#2) #3}
\def\prl#1#2#3{Phys. Rev. Lett. {\bf #1} (#2) #3}

\def\atmp#1#2#3{Adv. Theor. Math. Phys. {\bf #1} (#2) #3}
\def\jhep#1#2#3{J. High Energy Phys. {\bf #1} (#2) #3}
%
%
\lref\gauba{L. Alvarez-Gaum\'e and J. L. F. Barb\'on, ``Non-linear
vacuum phenomena in non-commutative QED,'' hep-th/0006209.}%

\lref\gmms{R. Gopakumar, J. Maldacena, S. Minwalla and A. Strominger,
``S-duality and noncommutative gauge theories,'' hep-th/0005048.}

\lref\sst{N. Seiberg, L. Susskind and N. Toumbas, ``Strings in
background electric field, space/time noncommutativity and a new
noncritical string theory,'' hep-th/0005040, \jhep{0006}{2000}{021}.}

\lref\br{J. L. F. Barb\'on and E. Rabinovici, ``Stringy fuzziness as the
custodian of time-space noncommutativity,'' hep-th/0005073.}

\lref\sstst{N. Seiberg, L. Susskind and N. Toumbas, ``Space/time
non-commutativity and causality,'' hep-th/0005015.}

\lref\gmss{R. Gopakumar, S. Minwalla, N. Seiberg and A. Strominger,
``(OM) theory in diverse dimensions,'' hep-th/0006062.}

\lref\km{I. R. Klebanov and J. Maldacena, ``$1+1$ dimensional NCOS and
its $U(N)$ gauge theory dual,'' hep-th/0006085.}

\lref\seiwit{N. Seiberg and E. Witten, ``String theory and
noncommutative geometry,'' hep-th/9908142, \jhep{9909}{1999}{032}.}

\lref\abs{O. Aharony, M. Berkooz and N. Seiberg, ``Light cone
description of $(2,0)$ superconformal theories in six dimensions,''
hep-th/9712117,
\atmp{2}{1998}{119}.}

\lref\bbss{E.~Bergshoeff, D.~S.~Berman, J.~P.~van der Schaar and
P.~Sundell, ``A noncommutative M-theory five-brane,'' hep-th/0005026.}

\lref\bbsstwo{E.~Bergshoeff, D.~S.~Berman, J.~P.~van der Schaar and
P.~Sundell, ``Critical fields on the M5-brane and noncommutative open
strings,'' hep-th/0006112.}

\lref\gm{J. Gomis and T. Mehen, ``Space-time noncommutative field
theories and unitarity,'' hep-th/0005129.}

\lref\natistr{N. Seiberg, ``New theories in six dimensions and matrix
description of M theory on $T^5$ and $T^5/Z_2$,'' hep-th/9705221,
\pl{408}{1997}{98}.} 

\lref\brs{M. Berkooz, M. Rozali and N. Seiberg, ``Matrix description of
M theory on $T^4$ and $T^5$,'' hep-th/9704089, \pl{408}{1997}{105}.}

\lref\abkss{O. Aharony, M. Berkooz, S. Kachru, N. Seiberg and
E. Silverstein, ``Matrix description of interacting theories in six
dimensions,'' hep-th/9707079, \atmp{1}{1998}{148}.}

\lref\edhiggs{E. Witten, ``On the conformal theory of the Higgs
branch,'' hep-th/9707093, \jhep{9707}{1997}{003}.}

\lref\sethi{S. Sethi, ``The matrix formulation of type IIB
five-branes,'' hep-th/9710005, \np{523}{1998}{158}.}

\lref\ganset{O. J. Ganor and S. Sethi, ``New perspectives on
Yang-Mills theories with sixteen supersymmetries,'' hep-th/9712071,
\jhep{9801}{1998}{007}.} 

\lref\matder{N. Seiberg, ``Why is the matrix model correct ?''
hep-th/9710009, \prl{79}{1997}{3577}.}

\lref\olddvv{R. Dijkgraaf, E. Verlinde and H. Verlinde, ``BPS
quantization of the fivebrane,'' hep-th/9604055, \np{486}{1997}{77};
``5d black holes and matrix strings,'' hep-th/9704018, \np{506}{1997}{121}.}

\lref\ab{O. Aharony and M. Berkooz, ``IR dynamics of $d=2$, ${\cal
N}=(4,4)$ gauge theories and DLCQ of `little string theories',''
hep-th/9909101, \jhep{9910}{1999}{030}.}

\lref\cds{A. Connes, M. R. Douglas and A. Schwarz, ``Noncommutative
geometry and matrix theory: compactification on tori,'' hep-th/9711162,
\jhep{9802}{1998}{003}.}%

\lref\dh{M. R. Douglas and C. Hull, ``D-branes and the noncommutative
torus,'' hep-th/9711165, \jhep{9802}{1998}{008}.}%

\lref\fradts{E. S. Fradkin and A. A. Tseytlin, ``Nonlinear
electrodynamics from quantized strings,'' \plb{163}{1985}{123}.}

\lref\clny{C. G. Callan, C. Lovelace, C. R. Nappi and S. A. Yost,
``String loop corrections to beta functions,'' \np{288}{1985}{525};
A. Abouelsaood, C. G. Callan, C. R. Nappi and S. A. Yost,
``Open strings in background gauge fields,'' \np{280}{1987}{599}.}%

\lref\chewu{G-H. Chen and Y-S. Wu, ``Comments on noncommutative open
string theory: V-duality and holography,'' hep-th/0006013.}%

\lref\har{T. Harmark, ``Supergravity and space-time non-commutative
open string theory,'' hep-th/0006023.}%

\lref\lursi{J. X. Lu, S. Roy and H. Singh, ``((F, D1), D3) bound state,
S-duality and noncommutative open string/Yang-Mills theory,''
hep-th/0006193.}%

\lref\jorjab{J. G. Russo and M. M. Sheikh-Jabbari, ``On noncommutative
open string theories,'' hep-th/0006202.}%

\lref\grs{O. J. Ganor, G. Rajesh and S. Sethi, ``Duality and
noncommutative gauge theory,'' hep-th/0005046.}%

\lref\nunossch{C. Nu\~nez, K. Olsen and R. Schiappa, ``From
Noncommutative Bosonization to S-Duality'', JHEP {\bf 0007} (2000)
030,  hep-th/0005059.}%

\myTitle{\vbox{\baselineskip12pt\hbox{hep-th/0006236}
\hbox{RUNHETC-2000-28}\hbox{CALT-68-2285}\hbox{CITUSC/00-038}}}
{\vbox{
\centerline{On Theories With Light-Like Noncommutativity}}}
\vskip 4pt
\centerline{Ofer Aharony$\,^{1}$, Jaume Gomis$\,^{2}$
 and Thomas
Mehen$\,^{2}$}
\medskip
\medskip
\medskip
\centerline{\it $^{1}$Department of Physics and Astronomy, Rutgers
University, 
Piscataway, NJ 08855}
\medskip
\centerline{\it $^{2}$ Department of Physics, California Institute of
Technology, Pasadena, CA 91125}   
\centerline{\it and}
\centerline{\it Caltech-USC Center for Theoretical Physics} 
\centerline{\it University of Southern California}
\centerline{\it Los Angeles, CA 90089}

\centerline{\tt oferah@physics.rutgers.edu; gomis,\  mehen@theory.caltech.edu}
\bigskip
\bigskip
\noindent

We show that field theories with light-like noncommutativity, that is
with $\theta^{0i}=-\theta^{1i}$,  are
unitary quantum theories, and that they can be obtained as decoupled
field theory limits of string theory with D-branes in a background
NS-NS $B$
field. For general noncommutativity parameters, we show that
noncommutative field theories which are unitary
can be obtained as decoupled field theory limits of string theory, while
those that are not unitary cannot be obtained from string theory
because massive open strings do not decouple.
We study the
different theories with light-like noncommutativity
which arise from Type II D-branes.
The decoupling limit of the D4-brane seems to lead to
a noncommutative field theory deformation of the $(2,0)$ SCFT of
M5-branes, while the D5-brane case leads to a noncommutative variation
of ``little string theories''. We discuss the DLCQ description of these
theories.

\medskip

\Date{June 2000}

\newsec{Introduction}

Theories on noncommutative spaces, in which the coordinates satisfy
$[x^\mu, x^\nu] = i \theta^{\mu \nu}$, have been a very active topic
of research in the last few years. They 
appear in decoupling limits of D-branes in string theory in
backgrounds with non-zero NS-NS $B$ fields \refs{\cds,\dh,\seiwit}.
The initial research focused on theories with only
space-like  
noncommutativity, that is with $\theta^{0i}=0$.  Gauge theories with
space-like noncommutativity 
arise from a decoupling limit of string theory involving D-branes with
non-zero space-like $B$ fields \seiwit, in which all string modes decouple and
one is left with a field theory (coming from the massless open
strings ending on the D-branes). Field theories on
such spaces are unitary.

Recently, it was realized that theories with time-like
noncommutativity, that is $\theta^{0i} \neq 0$, may also exist. However, 
field theories on such spaces exhibit acausal 
behaviour 
\refs{\sstst,\gauba}  
and the  quantum theories are not unitary
\gm. In \refs{\sst,\gmms,\br} 
it was found that a decoupled field theory limit for
D-branes with a  time-like $B$ field does not exist. However,
references \refs{\sst,\gmms} found a limit in
which the closed strings decouple but the massive open strings do not,
so this  limit describes
a noncommutative  open string theory (NCOS)
rather than a field 
theory. These open string theories were further analyzed in
\refs{\gmss,\km}. Several related aspects were recently considered in
\refs{\bbss,\bbsstwo,\chewu,\har,\lursi,\jorjab}.

In this paper we wish to analyze a third type of noncommutativity, in
which the noncommutativity parameter $\theta^{\mu \nu}$ is
light-like, for example with $\theta^{0i}=-\theta^{1i}$ (in light-cone
coordinates this corresponds to $\theta^{i-}\neq 0$). We will argue
that despite the
nonlocality in the time coordinate due to $\theta^{0i}\neq 0$, field
theories with light-like noncommutativity are quantum mechanically
unitary and exhibit interesting properties.

In section $2$ we determine which string backgrounds with a
constant $B$  field admit a decoupled noncommutative field theory  limit,
and verify for a
light-like $B$ (say, for $B_{0i}=B_{1i}\neq 0$) that such a field theory limit
exists. 
In section $3$ we analyze perturbative unitarity of noncommutative field
theories with arbitrary noncommutativity matrix $\theta^{\mu\nu}$. We
show that noncommutative 
field theories which  can be
obtained  as decoupled field theory limits of string theory are
perturbatively unitary quantum theories. On the other hand, those
noncommutative field theories that are not
unitary cannot be obtained from string
theory because massive open string modes do not decouple. 
Such theories can be made unitary by adding massive open string
degrees of freedom, decoupled from the closed strings, and lead to
NCOS theories.
The relation between unitarity in field theory and decoupling in
string theory is physically very appealing.

In section $4$ we analyze the decoupling limits of D-branes in Type II
string 
theory which lead to theories with light-like noncommutativity, 
decoupled from
closed strings and from massive open strings. 
In the case of light-like noncommutativity, the open string coupling
constant is identical to the closed string coupling constant. Therefore, the
analysis of the decoupling limits is completely analogous to the
analysis of decoupling limits of D-branes without a  $B$ field.
For D2-branes and D3-branes, we  find decoupling limits giving $2+1$
dimensional and $3+1$ dimensional super-Yang Mills (SYM) theories (with
light-like noncommutativity). 
The light-like noncommutative $3+1$ dimensional SYM theory
exhibits a conventional field theoretic S-duality, such that
the strong coupling limit of the noncommutative field
theory on the D3-brane is also a noncommutative ${\it field}$ theory
with light-like noncommutativity\foot{For a similar two-dimensional
phenomenon see 
\nunossch .}.
 For D4-branes we find that the decoupling limit seems to
lead to a $5+1$ dimensional field
theory (compactified on a circle), 
which is a noncommutative version of the $(2,0)$ six
dimensional SCFT\foot{Note that no such decoupled field theory exists for
space-like fields, since the self-duality of the 3-form on the 5-brane
forces a time-like noncommutativity to accompany any
space-like noncommutativity.}. 
For D5-branes we  find in the
decoupling limit a noncommutative version of ``little string
theories'', which reduces to $5+1$ dimensional noncommutative
SYM at low energies. Similar theories arise also from NS5-branes with
non-zero light-like RR backgrounds. For the various six dimensional
theories we also describe the discrete light-cone quantization (DLCQ) 
of the light-like
noncommutative theories, which is a simple variation of the DLCQ for
the same theories on a commutative space.

\newsec{Open Strings and Noncommutativity}

Consider open strings on a single D-brane (the generalization to
several overlapping D-branes is straightforward) in a constant background
electromagnetic 
field (or, equivalently, in a constant background NS-NS two-form field)
$B_{\mu\nu}$. The conformal field theory of this background was solved
in  
\refs{\fradts,\clny}. The dynamics of the open string is determined in
terms of the sigma model metric (closed string metric) $g_{\mu \nu}$, 
the background
two-form field $B_{\mu \nu}$ 
and the closed string coupling constant $g_s$. The
signature of space-time will be taken to be $(-,+,\cdots,+)$. 
The propagator of open string worldsheet coordinates between boundary
points $\tau $ and $\tau^\prime$ on the real axis of the upper half-plane
is\foot{Analogous expressions can be written for the
worldsheet superpartners $\psi^\mu$.}
\eqn\prop{
<X^\mu(\tau)\, X^\nu(\tau^\prime)>=-\alpha^\prime
G^{\mu\nu}\log(\tau-\tau^\prime)^2+{i\over
2}\theta^{\mu\nu}\hbox{sign}(\tau-\tau^\prime),}
where
\eqn\defini{\eqalign{
G^{\mu\nu}&=\left({1\over g+2\pi\alpha^\prime
B}\right)_S^{\mu\nu}=\left({1\over 
g+2\pi\alpha^\prime B}g{1\over
g-2\pi\alpha^\prime B}\right)^{\mu\nu},\cr 
\theta^{\mu\nu}&=2\pi\alpha^\prime\left({1\over g+2\pi\alpha^\prime
B}\right)_A^{\mu\nu}=-(2\pi \alpha^\prime)^2\left({1\over
g+2\pi\alpha^\prime B}B{1\over
g-2\pi\alpha^\prime B}\right)^{\mu\nu},}}
and the effective open string coupling is given by
\eqn\coupling{G_o = g_s \sqrt{{\det(g+2\pi \alpha^\prime B)}\over
\det(g)}.}

The classical effective action on the D-brane is obtained from the S-matrix
of open string states on the disc worldsheet. The
presence of the term proportional to $\theta^{\mu\nu}$ in the
propagator replaces the conventional product of fields in the effective
action with the $\star$-product of fields. 

We are interested in finding  
which electromagnetic backgrounds $B$ admit a decoupled field theory
limit such that the low energy effective description is given by a 
noncommutative field theory of the massless open string 
modes\foot{Clearly, there is always a low energy
limit whose description is given by conventional (commutative) field
theory. Here we 
are interested in a noncommutative field theory
description.}. Moreover, we want to determine which noncommutative
field theories are unitary quantum theories (see section $3$). We will
see that those four dimensional noncommutative field theories that are
perturbatively 
unitary are precisely those that can be obtained as a decoupled field
theory limit of string theory. 
Moreover, the noncommutative field theories
that are not unitary correspond to string backgrounds in which 
 the noncommutative massless open strings do not decouple from the
massive ones.

Our analysis will be based on looking at the Dirac-Born-Infeld action
describing constant electromagnetic background fields and seeing
when it describes a sensible theory on its own.
We  start by discussing the case of a D3-brane, for which we give a 
Lorentz-invariant description of the admissible backgrounds. 
Given a background $B_{\mu\nu}$ field, with particular values
for the electromagnetic Lorentz invariants\foot{We take $B_{0i}={\bf E}_i$
and $B_{ij}=\epsilon_{ijk}{\bf B}_k$.}
\eqn\loin{\eqalign{
I_1&={1\over 2}
B_{\mu\nu}B^{\mu\nu}={\bf B}^2-{\bf E}^2,\cr
I_2&={1\over 8}
\epsilon^{\mu\nu\rho\sigma}B_{\mu\nu}B_{\rho\sigma}={\bf E}\cdot
{\bf B},}}
one can perform a
Lorentz transformation to go to a standard frame where it is simple
to study the existence of a decoupled field theory limit. In the
standard frame, ${\bf E}$ can be chosen to be parallel, anti-parallel
or orthogonal to ${\bf B}$. There are 9
separate possibilities depending on $I_1$ and $I_2$. The standard
frames are  :
\medskip

\noindent
1) $I_1>0$ $I_2>0$ : ${\bf E}\| {\bf B}$,\ ${\bf B}^2>{\bf E}^2$;

\noindent
2) $I_1>0$ $I_2<0$ : $-{\bf E}\| {\bf B}$,\ ${\bf B}^2>{\bf E}^2$;

\noindent
3) $I_1<0$ $I_2>0$ : ${\bf E}\| {\bf B}$,\ ${\bf B}^2<{\bf E}^2$;

\noindent
4) $I_1<0$ $I_2<0$ : $-{\bf E}\| {\bf B}$,\ ${\bf B}^2<{\bf E}^2$;

\noindent
5) $I_1=0$ $I_2>0$ : ${\bf E}\| {\bf B}$,\ ${\bf B}^2={\bf E}^2$;

\noindent
6) $I_1=0$ $I_2<0$ : $-{\bf E}\| {\bf B}$,\ ${\bf B}^2={\bf E}^2$;

\noindent
7) $I_1>0$ $I_2=0$ : ${\bf E}\bot {\bf B}$,\ ${\bf B}^2>{\bf E}^2$;

\noindent
8) $I_1<0$ $I_2=0$ : ${\bf E}\bot {\bf B}$,\ ${\bf B}^2<{\bf E}^2$;

\noindent
9) $I_1=0$ $I_2=0$ : ${\bf E}\bot {\bf B}$,\ ${\bf B}^2={\bf E}^2.$
\medskip

It is known \seiwit\ that a space-like noncommutative field theory can
be obtained as  a decoupled 
limit of background 7), since one can always go to a frame in which only
the ${\bf B}$ field is non-zero. Moreover, background 8) can be boosted to a
frame in which only the ${\bf E}$ field is non-zero,
and \refs{\sst,\gmms,\br} showed that no
decoupled field theory limit exists for this background.
It is easy to see that whenever ${\bf E}$ is either parallel or
antiparallel to 
${\bf B}$ (backgrounds 1)-6))
there is no decoupled noncommmutative field theory
limit. The physical origin for the nonexistence of a decoupled field
theory limit is that in order to decouple the theory one must take
both ${\bf B}$ and  ${\bf E}$ large \seiwit , but whenever $I_2\neq 0$
there is 
an upper critical value of the electric field $E_c$ beyond which the
theory 
becomes unstable and, therefore, no sensible decoupled field
theory exists. In such a background the  ${\bf E}$ field reduces the
tension of a string when the string  is stretched in the
direction of ${\bf 
E}$, and it becomes tensionless precisely at $E_c$. Having a parallel
(anti-parallel)  ${\bf B}$ field does not change this phenomenon. 
More explicitly, consider the Dirac-Born-Infeld Lagrangian
density  for a
single D-brane in a background metric
$g_{\mu\nu}=\hbox{diag}(-g,g,g,g)$ and with arbitrary background ${\bf
B}$ and  ${\bf E}$ fields,
\eqn\bornin{
{\cal L}_{DBI}=-T_3\sqrt{-\hbox{det}(g_{\mu\nu}+2\pi\ap
B_{\mu\nu})}=-T_3\sqrt{g^4+(2\pi\ap)^2g^2({\bf 
B}^2-{\bf E}^2)-(2\pi\ap)^4({\bf E}\cdot{\bf
B})^2}.}
Clearly, whenever $I_2\neq 0$ the theory becomes unstable for
$|{\bf E}|>E_c\equiv g / (2\pi\ap)$ and, therefore, there is no decoupled
noncommutative field theory limit.

The only case left to consider is when $I_1=I_2=0$ (note that one
cannot always transform this case to ${\bf E}={\bf B}=0$, except by
an infinite Lorentz boost). This is the
light-like noncommutative case, where ${\bf E}^2={\bf B}^2$ and ${\bf
E}\cdot{\bf B}=0$. Clearly, there is no obstruction to taking the
decoupled field theory limit since there is no instability for any
value of the ${\bf
E}$ field. In this case, the presence of  the ${\bf B}$ field
perpendicular to  ${\bf E}$ forbids 
the ${\bf
E}$ field from reducing the energy of the string so that it becomes
tensionless.
Summarizing, the Lorentz invariant criterion for
backgrounds from 
which one can find a four dimensional decoupled field theory limit is
 $I_1\geq 0$ and $I_2=0$. The remaining backgrounds can be made
unitary by adding massive open string degrees of freedom, decoupled
from closed strings, and can lead to NCOS theories in an appropriate limit.
This criterion will be recovered in the
following section from a field theoretic analysis of unitarity.

Similarly,
it is easy to show for any Dp-brane with $p\geq2$ that a
light-like noncommutative field theory can also be obtained from string
theory in a background NS-NS $B$-field
$B_{0i}=B_{1i}$. For D2-branes the only Lorentz-invariant that can be
constructed from the background field is
 $I_1={1\over 2}B_{\mu\nu}B^{\mu\nu}$. The possible
cases are $I_1>0$ leading to the usual noncommutative Yang-Mills theory,
$I_1<0$ leading to the noncommutative open string theory, and $I_1=0$
which is the light-like case that we will discuss here.

\newsec{Unitarity Constraints}

In \gm\ unitarity of space-like noncommutative field theories and
time-like noncommutative theories was studied at the one loop
level, and it was found that space-like noncommutative theories are
unitary while time-like noncommutative theories are not unitary. One
can easily perform a general analysis of which types of
noncommutativity lead to unitary theories and which do not.

Unitarity requires \gm\ that the inner product $p\circ p$
is never negative, where $p$ is some external momentum and 
\eqn\pop{
p\circ p\equiv 
-p_\mu\theta^{\mu\rho}G_{\rho\sigma}\theta^{\sigma\nu}p_\nu\equiv p_\mu
g^{\mu\nu}_\theta p_\nu\geq 0,}
where $\theta^{\mu\nu}$ is the noncommutativity matrix and
$G_{\rho\sigma}$ is the background metric of the field theory.
The reason behind this requirement is that in order to define loop
integrals in these theories, one must analytically continue the
momentum and $\theta^{\mu\nu}$ to Euclidean space such that the
Euclidean expression for $p\circ
p$ is positive\foot{We will avoid values of the external momenta for
which $p\circ
p=0$, which lead to peculiar infrared divergences.}, so that
 Feynman graphs are well-defined. In order to check unitarity of the theory
one must analytically continue answers to Minkowski space. Therefore,
if in Minkowski space $p\circ p< 0$, Green's functions acquire
branch cuts as a function of momentum. It is the
presence of these extra branch cuts\foot{Green's functions in these
theories also have the conventional physical branch cuts associated with
threshold production of multiparticle states.}
 that causes nonunitary answers,
since they lead to extra imaginary pieces for S-matrix elements that
violate the optical theorem.

We will analyze in detail the four dimensional case and comment below
on the other cases.
A necessary condition for unitarity is that the eigenvalues of
$g^{\mu\nu}_\theta$ are nonnegative. This ensures that  $p\circ p\geq
0$ and that no unphysical branch cuts in Green's functions
appear. Therefore, we demand that 
\eqn\conda{
\hbox{det}(g^{\mu\nu}_\theta)=\hbox{det}(-\theta^{\mu\rho}G_{\rho\sigma}
\theta^{\sigma\nu})\geq 0.}
It is useful to rewrite the background metric of the field theory as 
\eqn\open{
G_{\mu\nu}=\left((g-2\pi\ap B)g^{-1}(g+2\pi\ap B)\right)_{\mu\nu}.}
Using \defini\ it follows that 
\eqn\met{
\hbox{det}(g^{\mu\nu}_\theta)=
(2\pi\ap)^{4}\hbox{det}\left(-{1\over g+2\pi\ap B}
Bg^{-1}B{1\over g-2\pi\ap B}\right).} 
Using the fact that $\hbox{det}(g+2\pi\ap B)=\hbox{det}(g-2\pi\ap B)$ one gets 
\eqn\metfin{
\hbox{det}(g^{\mu\nu}_\theta)=(2\pi\ap)^{4}{1\over \hbox{det}^2(g+2\pi\ap
B)\hbox{det}(-g)}\hbox{det}^2(B).}
Now, since $\hbox{det}(-g)<0$ and  $\hbox{det}^2(g+2\pi\ap B)\geq 0$, and
\eqn\detexp{
\hbox{det}^2(B)=({\bf E}\cdot {\bf B})^4=I_2^4,}
a necessary condition for unitarity is that
\eqn\unitco{
I_2={\bf E}\cdot {\bf B}=0.}

Therefore, there are three cases to be considered that can lead to a
unitary quantum field theory\foot{We will take the open string metric to
be $G^{\mu \nu} = \eta^{\mu \nu}$ in the equations below, a different
metric with the same signature will lead to the same results.} :

\noindent
7) In this case one can transform to a frame in which only the ${\bf
B}$-field is non-zero, for example $B_{12}\neq 0$. This leads to space-like
noncommutativity with 
$\theta^{12}=\theta$. Then, we have $p\circ p=\theta^2(p_1^2+p_2^2)\geq 0$
and the theory is unitary.

\noindent
8) In this case one can go to a frame in which only the ${\bf
E}$-field is non-zero,  for example $B_{01}\neq 0$. This leads to time-like
noncommutativity with 
$\theta^{01}=\theta$. Then, $p\circ p=\theta^2(p_0^2-p_1^2)$ 
can be negative and the theory is not unitary.

\noindent
9) In this case one can go to a frame with $B_{02}=B_{12}$. This leads
to light-like noncommutativity with
$\theta^{02}=-\theta^{12}=\theta$. 
Then, $p\circ
p=\theta^2(p_0-p_1)^2\geq 0$ and the theory is unitary.

Therefore, there is precise agreement between the backgrounds which have a
decoupled noncommutative field theory limit and the field theories which
have a perturbatively unitary S-matrix. It is easy to generalize this
also to other dimensions : the behavior of $p\circ p$ in the presence
of space-like, time-like and light-like noncommutativity is always as
in the cases 7), 8) and 9) discussed above (respectively).
In the rest of the paper we
will concentrate on theories with light-like noncommutativity.

\newsec{Decoupling Limit With Light-Like Noncommutativity}

In the previous two sections we showed that there could exist a
decoupled light-like noncommutative field theory limit of string
theory, and that the resulting 
field theory is quantum mechanically unitary. In this 
section we will study this decoupled field theory limit in detail for all
D-branes of Type II string theory.
It is convenient to analyze such decoupled field theories in 
light-cone coordinates,
$x^{\pm} = {1\over \sqrt{2}} (x^0 \pm x^1)$. 

By a Lorentz transformation we can
always choose the light-like noncommutativity parameter to be
$\theta^{2-} \equiv \Theta \neq 0$, 
with all other noncommutativity parameters
vanishing. In the usual coordinates such noncommutativity appears
whenever $\theta^{20}=-\theta^{21}=\Theta/\sqrt{2}$.
Such a configuration involves
noncommutativity in the time direction ($\theta^{20} \neq 0$), which
results in a theory non-local in time. 
Naively, one
would not expect such a theory to be unitary, nor would one expect that
it can be obtained from a decoupled limit of string theory.
 However, we
can always choose to 
perform a light-cone quantization in which $x^+$ is the time
coordinate. The field theory is local in the $x^+$ time
coordinate since $\theta^{i+}=0$. Therefore, one would expect the
light-cone Hamiltonian $H\equiv P_+$ to be 
Hermitean, 
and the field theory to be well-defined.
 In this section we 
describe how to get a field theory with this type of noncommutativity
as a limit of string theory.

We  start with $k$ D$p$-branes with general $p$, 
but we will focus only on the
first three coordinates\foot{In order to have a light-like noncommutative
field theory $p\geq 2$.} since the others will always have a flat
metric and no background fields. We take the closed string metric to be the
Minkowski metric $g_{\mu \nu} = \eta_{\mu \nu}$, and turn on a
non-zero $B_{2+}$. Using \defini\ \seiwit, we find the open string
metric  
$G^{+-} = -G^{22} = -1$, $G^{--} = -(2\pi \ap B_{2+})^2$, and the
noncommutativity parameter $\theta^{2-} = (2\pi \ap)^2 B_{2+}$. We
wish to discuss a decoupling limit in which we take $\ap \to 0$ to
decouple the closed strings and the massive open strings.
In order to obtain a finite noncommutativity parameter $\theta^{2-}
\equiv \Theta$ in the gauge theory
we need to take a very large $B_{2+} = \Theta / (2\pi
\ap)^2$.  Equivalently, one can turn on a constant flux in the overall $U(1)$
factor in the D-brane gauge group,
$F_{2+} = \Theta / (2\pi \ap)^2$ (times the
identity matrix). This
requires taking a very large electric field $E_2$. As discussed in
section $2$, when the
background flux is light-like, 
a large electric field does not lead to an instability.

At
first sight, we end up in this limit with a strange open string metric
with an infinite
$G^{--}$ component. However, this does not actually have any physical
effect, 
and we can easily  fix this\foot{This was
suggested to us by N. Seiberg.} by a change of coordinates  
\eqn\newcoor{y^+ \equiv x^+;\qquad y^- \equiv x^- + {1\over 2} G^{--}
x^+;\qquad y^i \equiv x^i\ (i=2,\cdots,p).} 
In the new coordinates the open
string metric is $G^{\mu \nu} = \eta^{\mu \nu}$ and we have a finite
noncommutativity parameter $\theta^{2-}=\Theta$, so we obtain precisely the
field theories discussed above.
Equivalently, we could have started with a closed string metric with
$g_{++}$ which goes to infinity such that the open string metric is
diagonal; this situation is related to the situation we describe here
by a shift similar to \newcoor.

It is important to note that the theories with light-like
noncommutativity which we discuss 
here do not have a typical noncommutativity scale in them, since
there is no Lorentz-invariant scalar one can make out of
$\theta^{2-}$. Longitudinal Lorentz boosts can rescale $\theta^{2-}$
to any 
(non-zero) 
value we wish it to be. The scaling of $B_{2+}$ which we describe
above is the one which gives $\theta^{2-}=\Theta=constant$ in the decoupling
limit, but any scaling of these parameters (which gives a
non-zero and finite $\theta^{2-}$) 
is related by a boost to the scaling we describe above.
Correlation functions in these theories
depend on the longitudinal boost invariant
combination $\theta^{2-} P_-$.

Using \coupling\ we find that
the open string coupling constant in this case is the same as it
was without the $B$ field, namely $G_o = g_s$, so that the
Yang-Mills (YM) coupling constant is given by the usual formula $g_{YM}^2 =
(2\pi)^{p-2} g_s (\ap)^{(p-3)/2}$. The discussion of the possible
decoupling limits is thus exactly the same as without the $B$ field
and not the same as in the case of a space-like $B$ field.
One scales $\ap \to 0$ to decouple the field theory from the bulk and
scales $g_s$ such that one is left with a non-trivial field theory on
the brane ($g_{YM}$ is kept fixed).
We will now analyze the decoupled theories that we
get in different dimensions:

\subsec{D3-branes}

For $k$ D3-branes we can take $\ap \to 0$ keeping
$g_s$ fixed, and we get a $U(k)$ NCYM theory with finite
noncommutativity, which is decoupled from the closed strings and from
the massive open strings by the same arguments used in the absence of
the $B$ field. 

It is interesting to 
note that in the $3+1$ dimensional case the $U(k)$ gauge theories that
we find go to themselves under S-duality, unlike the
theories with space-like noncommutativity which are S-dual to
noncommutative open string theories (NCOS) \gmms. In the light-like
case the $3+1$ 
dimensional decoupled theories inherit the S-duality transformation
from Type IIB string theory. This transformation inverts the (complexified)
gauge coupling and changes the background flux; to leading order in
the background flux it exchanges $F_{\mu \nu}$ with
$(*F)_{\mu \nu}$ \grs, where $*$ denotes the Hodge operation, and for
light-like fields this is actually the exact transformation. This leads
to a field theory with a light-like noncommutativity parameter
$\theta^{3-}$. Generally,
S-duality changes the light-like noncommutativity parameter by
$\theta^{i-} \to \epsilon^{ij} \theta^{j-}$, where the epsilon symbol
involves the directions transverse to the light-cone coordinates.

\subsec{D2-branes}

For $k$ D2-branes, if we want to keep the YM coupling
constant fixed as we take $\ap \to 0$ we must also scale $g_s \propto
(\ap)^{1/2} \to 0$ at the same time, but this obviously does not
affect the decoupling arguments. In this limit we find precisely
the $2+1$ dimensional $U(k)$ light-like noncommutative supersymmetric
gauge theory.

\subsec{D4-branes}

Things become more interesting if we discuss the decoupling limit for
$k$ D4-branes. In this case, if we wish to take $\ap \to 0$ and keep the
YM coupling constant fixed, we must scale $g_s$ to infinity as
$(\ap)^{-1/2}$. Thus, it is more appropriate to think of the theory as
M theory compactified on a circle. The Planck scale in M theory scales
as $M_p^3 = M_s^3 / g_s \propto (\ap)^{-1}$ so it goes to
infinity, while the radius of the M theory circle remains finite (as
in the absence of the $B$ field), $R_{11} = g_s (\ap)^{1/2} \simeq 
g_{YM}^2$. The
decoupled theory on the D4-branes should thus be viewed as a decoupled
theory on $k$ M5-branes compactified on a finite circle. This is not
surprising since the $4+1$ dimensional gauge theory on its own is
non-renormalizable even before we add the noncommutativity. 

When we go to M theory it is natural to keep the metric on the brane (which
is the same as the metric in the bulk up to an infinite $g_{++}$ which we
discussed above) in the form $G^{\mu \nu} = \eta^{\mu \nu}$. In these
coordinates the $x^{11}$ direction has periodicity $2\pi R_{11}$.
Translating the relation $B_{2+}
\simeq \Theta / (2\pi \ap)^2$ to M theory variables, we find that
the 3-form of M theory scales as
\eqn\oldcfield{C_{2(11)+} \simeq -\Theta R_{11} M_p^6}
in the limit we are taking, with $R_{11}$ constant and $M_p$ going to
infinity.

We claim that this limit,
for $k$ M5-branes
oriented in the $(0,1,2,3,4,11)$ directions, 
defines a decoupled ``noncommutative'' variation of the $(2,0)$
theory living on the M5-branes. 
In M theory, one can
gauge away any constant components of the background  $C$ field that
are transverse to 
the M5-branes, as well as the anti-self-dual components of $C$ along the
M5-branes. So, we can take the background $C$ field to be self-dual, with
nonvanishing $C_{34+} =
C_{2(11)+}$.
Equivalently, instead of the $C$ field
we can take the self-dual 3-form worldvolume field 
$H_{34+}=H_{2(11)+}$ on the M5-branes to scale in the same way that we
scaled the $C$ field in the
decoupling limit. Here, we used the fact that 
for light-like fields the non-linear
self-duality condition on the 3-form field $H$ actually becomes linear
\seiwit.

It is not clear how to characterize the ``noncommutativity'' (or whatever
generalizes this notion)
in the six dimensional theory. It seems reasonable to expect that this
theory has a 3-form ``generalized noncommutativity parameter'', which
would be (for example) the coefficient of the leading (dimension 9)
irrelevant operator appearing in the low-energy expansion of the theory.
If we call this parameter $\psi^{\mu \nu \rho}$, dimensional analysis
and Lorentz covariance
determine that in the light-like ``noncommutative'' case described above
it will be given
by $\psi^{2(11)-} = -C_{2(11)+} / M_p^6 \simeq \Theta R_{11}$.
This means if we take the $R_{11} \to \infty$ (or $g_{YM} \to \infty$)
limit in the theory described above, we
do not get a theory with finite ``noncommutativity''. Rather,
such a theory would arise from taking $C_{2(11)+} \simeq -\psi M_p^6$
with $\psi$ kept constant as $M_p \to \infty$.
However, since we do not understand the notion of ``generalized
noncommutativity'' we cannot rigorously justify these claims.
In \refs{\bbss,\bbsstwo} 
it was suggested that six dimensional ``noncommutative''
theories can be characterized by an open membrane metric which could be
analogous to
the open string metric described above; in our case 
this ``open
membrane'' metric turns out to be $\eta^{\mu\nu}$,
just like the open string metric on the D4-brane. The fact that the
``open membrane metric'' remains finite as we take $M_p$ to infinity
is consistent with our claim the the six dimensional theory is a
field theory, with no additional open strings or membranes.

A theory which seems to describe the DLCQ of the six dimensional
theory described above was discussed in section 4 of
\abs. The decoupled theory of $k$ M5-branes with $N$ units of light-like
momentum ($P_- = N/R$) was described
in terms of the $g_{YM} \to \infty$ limit of the Higgs branch of the
${\cal N}=8$ $U(N)$
$0+1$ dimensional
SYM theory with $k$ hypermultiplets in the fundamental representation \abkss,
and the Fayet-Iliopoulos (FI) 
parameters of this theory were identified (in a particular
normalization) with $C_{ij+} / R
M_p^6$ (where $R$ is the radius of the compact light-like
direction). Note that in the DLCQ, where the $x^-$ direction is
compact, we can no longer perform arbitrary longitudinal
Lorentz boosts since these also
rescale the radius $R$; the combination $C_{ij+}/R$ appearing in the
DLCQ is boost-invariant and can thus be used to characterize the
``noncommutativity'' of the theory.
The fact that the DLCQ depends on $C_{ij+} / M_p^6$ is consistent with
our conjecture for the ``generalized noncommutativity parameter''
described above.
The infinite shift we found \newcoor\ between the closed string and
open string coordinates can be identified with
the infinite shift found in
the DLCQ between the vacuum energies of the Higgs and Coulomb branches 
(in the decoupling limit).

The relation between the six dimensional theory we described here and
the ``open membrane''
theories discussed in \refs{\gmss,\bbss,\bbsstwo} is not clear. Those
theories involve additional degrees of freedom  in addition to the six
dimensional field theory, while such 
degrees of freedom do not seem to appear in our case.

\subsec{D5-branes and NS5-branes}

For $k$ D5-branes, again we have to take $g_s$ to infinity as we take
$\ap$ to zero, in order to keep the Yang-Mills coupling fixed. The
strong coupling limit of type IIB string theory is described by the
S-dual theory, in which the string coupling goes to zero. Thus, it is
best to describe the limit we are discussing in the S-dual theory. In
this theory we find that we have $k$ NS 5-branes, the string coupling
goes 
to zero, and the string tension (which is the inverse gauge coupling
on the NS5-branes) remains constant. This is the same limit used to
define ``little string theories'' (LSTs) \refs{\natistr,\brs,\olddvv}, 
so the theory we  get in
this limit is a noncommutative version of the LSTs. The
S-duality turns the NS-NS $B$ field into a RR
B-field. Therefore,
we are 
discussing NS 5-branes with a constant RR $B_{2+}$ field which
goes to 
infinity. Equivalently (as in the previous cases) we can just take the
gauge field strength $F_{2+}$ on the 5-branes to go to infinity. At low
energies (compared to the string tension)
this limit gives a (non-renormalizable)
light-like noncommutative $5+1$ dimensional
gauge theory, while at energies of the order of the string scale we
have the full 
noncommutative LST.

As in the previous discussion, the DLCQ of this NCLST is given by a
simple deformation of the DLCQ of the LST with $(1,1)$ supersymmetry
\refs{\sethi,\ganset}. 
This DLCQ description (which is reviewed in \ab), 
for the theory of $k$ 5-branes with $N$ units of
light-like momentum, is given by the low-energy SCFT of the Coulomb
branch of the $1+1$ dimensional $U(N)^k$
gauge theory with bifundamental hypermultiplets
for consecutive $U(N)$ groups (arranged in a circle). The
noncommutative deformation is realized in the DLCQ by adding an equal mass
to the $k$ bifundamental hypermultipets. Note that this mass, like the
light-like noncommutative parameter, is a vector of the $SO(4)$
rotation group
acting on the four transverse coordinates of the 5-branes.

A similar deformation exists also for the $(2,0)$-supersymmetric LST
arising from NS 5-branes in type IIA string theory. The
``noncommutative'' deformation now involves a constant RR 3-form field
$C_{ij+}$, or equivalently a constant 3-form field in the 5-brane
worldvolume. In the DLCQ this deformation corresponds (as discussed in
\ab) to turning on a
Fayet-Iliopoulos term in the corresponding $1+1$ dimensional gauge
theory \refs{\abkss,\edhiggs}. At low energies (compared to the string scale)
the $(2,0)$ ``noncommutative'' LST 
reduces to the $(2,0)$ ``noncommutative'' field theory
arising from $k$ M5-branes, which we described in the previous subsection.

For higher dimensional D-branes there seems to be no decoupling limit
from the bulk, just like in the case without the noncommutativity.

\vskip 1cm
{\bf Acknowledgements:}
We would like to thank M. Berkooz and especially N. Seiberg for useful
discussions. The work of O.A. was supported in part by DOE grant
DE-FG02-96ER40559. The work of J.G. and T.M. is supported in part by
the DOE under grant no. DE-FG03-92-ER 40701.
O.A. would like to thank the USC/Caltech Center
for Theoretical Physics and the Erwin Schr\"odinger Institute in
Vienna for hospitality during the course of this work. J.G. would like
to thank CERN for hospitality during part of this work.

\listrefs

\end